\documentclass[10pt,conference]{IEEEtran}
\IEEEoverridecommandlockouts
\usepackage{booktabs}
\usepackage[utf8]{inputenc}
\usepackage{amsmath,amsfonts}
\usepackage{array}
\usepackage[caption=false,font=scriptsize,labelfont=rm,textfont=rm]{subfig}
\usepackage{textcomp}
\usepackage{stfloats}
\usepackage{url}
\usepackage{verbatim}
\usepackage{graphicx}
\usepackage{cite}
\usepackage{CJKutf8}
\usepackage{amsthm,amsmath,amssymb} 
\usepackage{algpseudocode}   
\usepackage{mathrsfs}
\usepackage{multicol}
\usepackage{svg}
\usepackage{graphicx}
\usepackage[ruled,linesnumbered]{algorithm2e}
\usepackage{algpseudocode}
\usepackage[colorlinks,linkcolor=blue,anchorcolor=blue,citecolor=blue]{hyperref}

\begin{document}

\title{SIM-Assisted End-to-End Co-Frequency Co-Time Full-Duplex System
\thanks{This paper was partially funded by the National Key R \& D
Program of China (2020YFB1806602), BUPT-China Unicom Joint Innovation Center and Fundamental Research Funds for the Central Universities (2242022k60006). \textit{(Corresponding author: Qiang Wang, Qiuyan Liu.)}
}
\thanks{Simulation code:}
}

 \author{\IEEEauthorblockN{Yida Zhang\IEEEauthorrefmark{1}, Qiuyan Liu\IEEEauthorrefmark{2}, Yuqi Xia\IEEEauthorrefmark{1}, Guoxu Xia\IEEEauthorrefmark{1}, Qiang Wang\IEEEauthorrefmark{1},}
 
 \IEEEauthorblockA{\IEEEauthorrefmark{1}
National Engineering Research Center for Mobile Network Technologies,\\
Beijing University of Posts and Telecommunications, Beijing 100876, China, \\
\IEEEauthorrefmark{2} China United Network Communications Corporation Research Institute, Beijing 100037, China\\
Email: \{zhangyida02, xiayuqi, xgx797, wangq\}@bupt.edu.cn, liuqy95@chinaunicom.cn}
}

\maketitle

\begin{abstract}
To further suppress the inherent self-interference (SI) in co-frequency and co-time full-duplex (CCFD) systems, we propose integrating a stacked intelligent metasurface (SIM) into the RF front-end to enhance signal processing in the wave domain. Furthermore, an end-to-end (E2E) learning-based signal processing method is adopted to control the metasurface. Specifically, the real metasurface is abstracted as hidden layers of a network, thereby constructing an electromagnetic neural network (EMNN) to enable driving control of the real communication system. Traditional communication tasks, such as channel coding, modulation, precoding, combining, demodulation, and channel decoding, are synchronously carried out during the electromagnetic (EM) forward propagation through the metasurface. Simulation results show that, benefiting from the additional wave-domain processing capability of the SIM, the SIM-assisted CCFD system achieves significantly reduced bit error rate (BER) compared with conventional CCFD systems. Our study fully demonstrates the potential applications of EMNN and SIM-assisted E2E CCFD systems in next-generation transceiver design.
\end{abstract}

\begin{IEEEkeywords}
Stacked intelligent metasurfaces (SIM), Co-frequency co-time
full-duplex (CCFD), electromagnetic neural network (EMNN), End-to-end (E2E).
\end{IEEEkeywords}

\section{Introduction.} 
Co-frequency and co-time full duplex (CCFD) has attracted widespread attention due to its potential to achieve double spectral efficiency by enabling simultaneous information exchange within the same frequency band~\cite{7024120}. However, the implementation of this technology still faces significant challenges, constrained by the inherent self-interference (SI) between the transmit and receive antennas of the same terminal. In response, traditional approaches typically employ passive isolation techniques to mitigate SI by separating distance, direction, and polarization~\cite{6353396,9841615}.

Recently, stacked intelligent metasurfaces (SIMs) have emerged as a key technology, representing three-dimensional metasurface devices that utilize cascaded metasurface layers to perform electromagnetic (EM) wave-based analog computing~\cite{liu2022programmable}. In practice, SIMs are often integrated into the radio-frequency (RF) front end to regulate the amplitude and phase of transmitted or received signals, thereby assisting signal processing tasks in the wave domain. This integrated architecture is referred to as holographic multiple-input multiple-output (HMIMO)~\cite{10819473}. It effectively reduces the required number of RF chains while relying on its wave-domain information processing capability to achieve better suppression of SI~\cite{11091527}.

While hardware architectures continue to evolve, end-to-end (E2E) learning has gradually become an important approach for enhancing the performance of next-generation wireless systems~\cite{8214233}. Unlike traditional modular optimization, E2E learning deploys neural networks at both the transmitter and receiver to jointly optimize processes such as coding, modulation, beamforming, and detection, thereby approaching global optimal performance under hardware constraints~\cite{8054694}. For SIM-assisted HMIMO transceivers, E2E learning can adaptively configure parameters in both the digital and wave domains, fully exploiting the potential of metasurfaces while maintaining robustness under hardware imperfections. Consequently, it is regarded as a key direction for intelligent 6G transceiver design~\cite{8985539}.

Inspired by the additional signal processing capability that SIM can exhibit in the wave domain, which is expected to bring new degrees of freedom for SI suppression in CCFD systems, we propose a SIM-assisted E2E CCFD system and design an electromagnetic neural network (EMNN) for optimization. The main contributions of this paper are as follows:
\begin{enumerate}[]
    \item 
    We are the first to propose employing SIMs to assist signal processing in CCFD systems and present the corresponding mathematical model. The proposed system aims to achieve joint SI cancellation in both the digital and wave domains through the deployment of deep neural networks (DNNs) and the integration of SIM devices. 

    \item 
    To achieve overall performance optimization, we design an EMNN for the CCFD system by abstracting the real metasurface as hidden layers of a network. The EMNN links the parameters of the hidden layers to the EM units, thereby enabling model-driven control. 

    \item 
    We adopt transfer learning to accelerate the training process of the EMNN. Finally, a comprehensive evaluation of the performance of the SIM-assisted E2E CCFD system is conducted and compared with that of the conventional CCFD system.
    
\end{enumerate}

\section{System Model.} \label{II}

\begin{figure}[!t]
\centering
\includegraphics[width=2.7in]{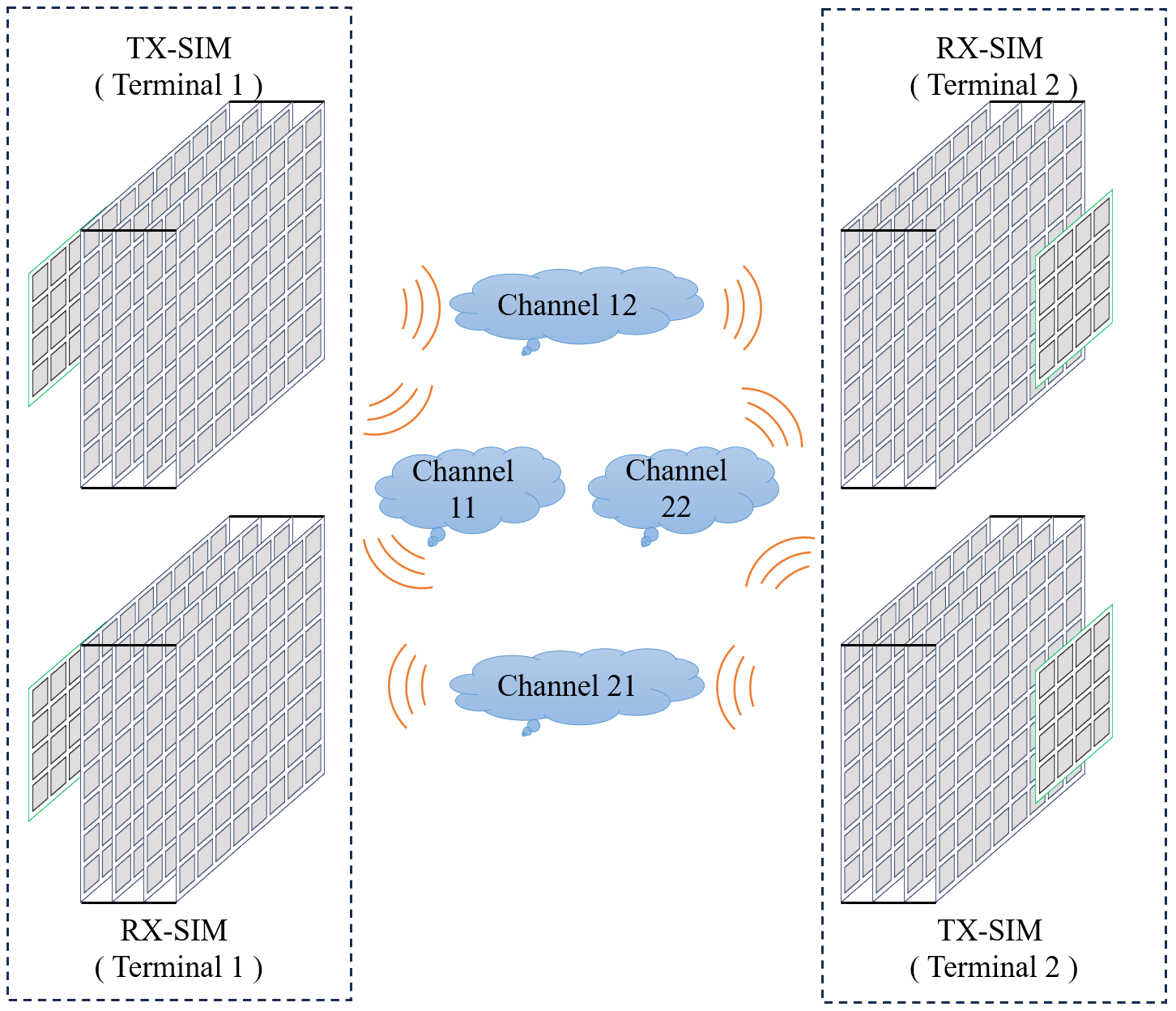}
\caption{Schematic diagram of the SIM-assisted CCFD system.}
\label{Fig1}
\vspace{-0.3cm}
\end{figure}

\begin{figure}[!t]
\centering
\includegraphics[width=2.8in]{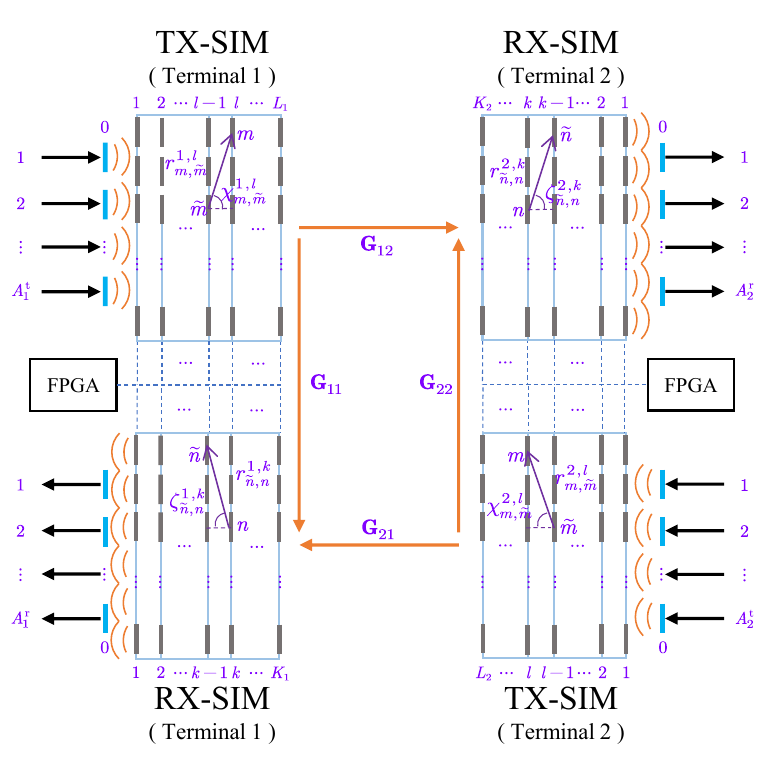}
\caption{SIM-assisted E2E CCFD system parameter diagram.}
\label{Fig2}
\end{figure}

\begin{figure*}[!b]
\vspace{-0.7cm}
\begin{align*}
&\overline{\ \ \ \ \ \ \ \ \ \ \ \ \ \ \ \ \ \ \ \ \ \ \ \ \ \ \ \ \ \ \ \ \ \ \ \ \ \ \ \ \ \ \ \ \ \ \ \ \ \ \ \ \ \ \ \ \ \ \ \ \ \ \ \ \ \ \ \ \ \ \ \ \ \ \ \ \ \ \ \ \ \ \ \ \ \ \ \ \ \ \ \ \ \ \ \ \ \ \ \ \ \ \ \ \ \ \ \ \ \ \ \ \ \ \ \ \ \ \ \ \ \ \ \ \ \ \ \ \ \ \ \ \ \ \ \ \ \ \ \ \ \ \ \ \ \ \ }\\
\label{eq1}
&\ \ \ \ \ \ \ \ \ \ \ \ \ \ \ \ \mathbf{\Phi }_{q}^{l}=\mathrm{diag}\left( e^{\mathrm{j}\theta _{1}^{q,l}},e^{\mathrm{j}\theta _{2}^{q,l}},\cdots ,e^{\mathrm{j}\theta _{M_q}^{q,l}} \right) \in \mathbb{C} ^{M_q\times M_q},\ \theta _{m}^{q,l}\in \left[ 0,2\pi \right) ,\ m\in \mathcal{M} _q,\ l\in \mathcal{L} _q,\ q\in \left\{ 1,2 \right\},
\tag{1} \\
\label{eq2}
&\ \ \ \ \ \ \ \ \ \ \ \ \ \ \ \ \mathbf{\Psi }_{q}^{k}=\mathrm{diag}\left( e^{\mathrm{j}\xi _{1}^{q,k}},e^{\mathrm{j}\xi _{2}^{q,k}},\cdots ,e^{\mathrm{j}\xi _{N_q}^{q,k}} \right) \in \mathbb{C} ^{N_q\times N_q},\ \xi _{n}^{q,k}\in \left[ 0,2\pi \right) ,\ n\in \mathcal{N} _q,\ k\in \mathcal{K} _q,\ q\in \left\{ 1,2 \right\}.\tag{2}\\
&\overline{\ \ \ \ \ \ \ \ \ \ \ \ \ \ \ \ \ \ \ \ \ \ \ \ \ \ \ \ \ \ \ \ \ \ \ \ \ \ \ \ \ \ \ \ \ \ \ \ \ \ \ \ \ \ \ \ \ \ \ \ \ \ \ \ \ \ \ \ \ \ \ \ \ \ \ \ \ \ \ \ \ \ \ \ \ \ \ \ \ \ \ \ \ \ \ \ \ \ \ \ \ \ \ \ \ \ \ \ \ \ \ \ \ \ \ \ \ \ \ \ \ \ \ \ \ \ \ \ \ \ \ \ \ \ \ \ \ \ \ \ \ \ \ \ \ \ \ }\\
\label{eq3}
&\ \ \ \ \ \ \ \ \ \ \ \ \ \ \ \ \left[ \mathbf{V}_{q}^{l} \right] _{m,\tilde{m}}=\frac{S\cos \chi _{m,\tilde{m}}^{q,l}}{r_{m,\tilde{m}}^{q,l}}\left( \frac{1}{2\pi r_{m,\tilde{m}}^{q,l}}-\mathrm{j}\frac{f}{c} \right) e^{\mathrm{j}2\pi r_{m,\tilde{m}}^{q,l}f/c},\ m\in \mathcal{M} _q,\ \tilde{m}\in \mathcal{M} _q,\ l\in \mathcal{L} _q,\ q\in \left\{ 1,2 \right\}, \tag{3}\\
\label{eq4}
&\ \ \ \ \ \ \ \ \ \ \ \ \ \ \ \ \left[ \mathbf{U}_{q}^{k} \right] _{\tilde{n},n}=\frac{S\cos \zeta _{\tilde{n},n}^{q,k}}{r_{\tilde{n},n}^{q,k}}\left( \frac{1}{2\pi r_{\tilde{n},n}^{q,k}}-\mathrm{j}\frac{f}{c} \right) e^{\mathrm{j}2\pi r_{\tilde{n},n}^{q,k}f/c},\ \tilde{n}\in \mathcal{N} _q,\ n\in \mathcal{N} _q,\ k\in \mathcal{K} _q,\ q\in \left\{ 1,2 \right\}.\tag{4}
\end{align*}
\end{figure*}

As shown in Fig.~\ref{Fig1}, we consider an E2E CCFD system comprising Terminal~1 and Terminal~2, where both the TX-SIM and RX-SIM are integrated within each terminal. $L_q$ and $K_q$ denote the numbers of metasurface layers for the TX-SIM and RX-SIM of terminal $q$, respectively, where $q \in \{1,2\}$. For notational convenience, we define the set of TX metasurfaces at terminal $q$ as $\mathcal{L}_q = \{0,1,2,\cdots,L_q\}$, with $M_q=M_{q}^{\mathrm{x}}\times M_{q}^{\mathrm{y}}$ EM units on each layer, and the set of RX metasurfaces as $\mathcal{K}_q = \{0,1,2,\cdots,K_q\}$, with $N_q=N_{q}^{\mathrm{x}}\times N_{q}^{\mathrm{y}}$ EM units on each layer, where $0 \in \mathcal{L}_q$ and $0 \in \mathcal{K}_q$ respectively denote the TX antenna metasurface and the RX antenna metasurface of the corresponding terminal. Moreover, $A_{q}^{\mathrm{t}} = A_{q}^{\mathrm{t}^{\mathrm{x}}} \times A_{q}^{\mathrm{t}^{\mathrm{y}}}$ and $A_{q}^{\mathrm{r}} = A_{q}^{\mathrm{r}^{\mathrm{x}}} \times A_{q}^{\mathrm{r}^{\mathrm{y}}}$ represent the numbers of TX antennas and RX antennas at terminal $q$, respectively. We assume that all metasurface EM units have identical dimensions, and the spacing between adjacent EM units is $d$.

The transmission coefficient matrices of the $l$-th layer for TX-SIM and the $k$-th layer for the RX-SIM at terminal $q$ are given by \eqref{eq1} and \eqref{eq2}, respectively. According to the Rayleigh-Sommerfeld diffraction theory~\cite{Lin2018Alloptical}, the transmission coefficient from the $\tilde{m}$-th EM unit on the $(l-1)$-th transmit metasurface layer to the $m$-th EM unit on the $l$-th transmit metasurface layer at terminal $q$ is expressed by~\eqref{eq3}, where $r_{m,\tilde{m}}^{q,l}$ denotes the corresponding transmission distance, $S=d\times d$ is the area of each EM unit in the SIM, while $\chi_{m,\tilde{m}}^{q,l}$ represents the angle between the propagation direction and the normal direction of the $(l-1)$-th transmit metasurface layer. Similarly, the transmission coefficient from the $n$-th EM unit on the $k$-th receive metasurface layer to the $\tilde{n}$-th EM unit on the $(k-1)$-th receive metasurface layer at terminal $q$ is expressed by~\eqref{eq4}, where $r_{\tilde{n},n}^{q,k}$ denotes the corresponding transmission distance, while $\zeta_{\tilde{n},n}^{q,k}$ represents the angle between the propagation direction and the normal direction of the $k$-th receive metasurface layer.

The cumulative effect generated by the layer-by-layer propagation of signals within SIM can be characterized by a chain of matrix products. For the TX-SIM at terminal $q$, the propagation coefficient matrix is
\begin{align*}
    \mathbf{T}_q=\mathbf{\Phi }_{q}^{L}\mathbf{V}_{q}^{L}\cdots \mathbf{\Phi }_{q}^{2}\mathbf{V}_{q}^{2}\mathbf{\Phi }_{q}^{1}\mathbf{V}_{q}^{1}\in \mathbb{C} ^{M_q\times A_{q}^{\mathrm{t}}},\ q\in \left\{ 1,2 \right\}.\tag{5}
\end{align*}
Similarly, for the RX-SIM at terminal $q$, the propagation coefficient matrix is
\begin{align*}
    \mathbf{R}_q=\mathbf{U}_{q}^{1}\mathbf{\Psi }_{q}^{1}\mathbf{U}_{q}^{2}\mathbf{\Psi }_{q}^{2}\cdots \mathbf{U}_{q}^{K}\mathbf{\Psi }_{q}^{K}\in \mathbb{C} ^{A_{q}^{\mathrm{r}}\times N_q},\ q\in \left\{ 1,2 \right\}.\tag{6}
\end{align*}

In summary, the signal received at terminal $q$ is given by
\begin{gather*}
\label{eq1}
\mathbf{y}_q=\mathbf{R}_q\left( \mathbf{G}_{pq}\mathbf{T}_p\mathbf{p}_p\odot \mathbf{x}_p+\mathbf{G}_{qq}\mathbf{T}_q\mathbf{p}_q\odot \mathbf{x}_q \right) +\mathbf{n}_q,\\
p,q\in \{1,2\},\ p\ne q, \tag{7}
\end{gather*}
where $\mathbf{n}_q\in \mathbb{C} ^{A_{q}^{\mathrm{r}}\times 1}$ is the receiver noise vector with distribution $\mathcal{C} \mathcal{N} \left( 0,\sigma ^2\mathbf{I}_{A_{q}^{\mathrm{r}}} \right) $, $\mathbf{p}_q\in \mathbb{C} ^{A_{q}^{\mathrm{t}}\times 1}$ denote the vector of transmit powers at terminal $q$, $\mathbf{G}_{pq}$ is the channel coefficient matrix from the $L_p$-th layer metasurface of the TX-SIM at terminal $p$ to the $K_q$-th layer metasurface of the RX-SIM at terminal $q$, $\mathbf{x}_q\in \mathbb{C} ^{A_{q}^{\mathrm{t}}\times 1}$ is the signal vector satisfying $\mathbb{E} \left\{ \mathbf{x}_q\left( \mathbf{x}_q \right) ^{\mathrm{H}} \right\} =\mathbf{I}_{A_{q}^{\mathrm{t}}}$.

\begin{figure*}[!t]
\centering
\includegraphics[width=5in]{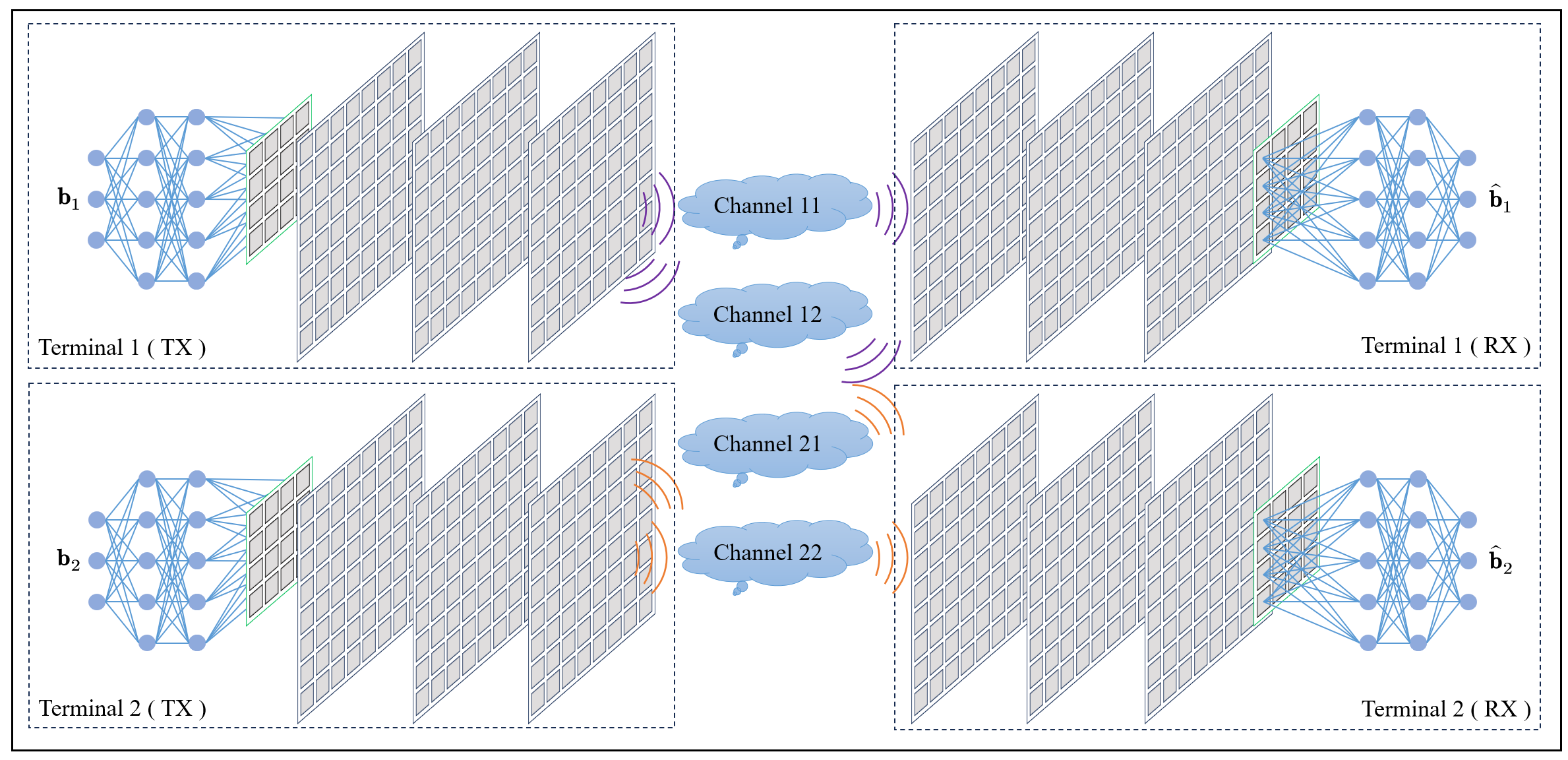}
\caption{Schematic diagram of the E2E CCFD system assisted by SIM.}
\label{Fig3}
\vspace{-0.4cm}
\end{figure*}

Let $N^{\mathrm{bit}}_p$ denote the number of bits corresponding to an symbol transmitted by the terminal $p$, $\mathbf{b}_p\in \{0,1\}^{N_{p}^{\mathrm{bit}}\times 1}$ and $\hat{\mathbf{b}}_q\in \{0,1\}^{N_{p}^{\mathrm{bit}}\times 1}$ denote the transmitted bit vector at terminal $p$ and the received bit vector at terminal $q$, respectively. Then, the bit transmission model of SIM-assisted E2E CCFD system model can be expressed as
\begin{gather*}
\hat{\mathbf{b}}_q=\mathbf{f}_{q}^{\mathrm{D}}\left( \mathbf{R}_q\left( \mathbf{G}_{pq}\mathbf{T}_p\mathbf{p}_p\odot \mathbf{x}_p+\mathbf{G}_{qq}\mathbf{T}_q\mathbf{p}_q\odot \mathbf{x}_q \right) +\mathbf{n}_q \right),
\\
p,q\in \{0,1\}, p\ne q. \tag{8}
\end{gather*}
where $\mathbf{x}_q=\mathbf{f}_{q}^{\mathrm{M}}(\mathbf{b}_q)$ denotes the modulation mapping from $\mathbf{b}_q$ to the signal at terminal $q$. Furthermore, $\hat{\mathbf{b}}_q=\mathbf{f}_{q}^{\mathrm{D}}\left( \mathbf{y}_q \right)  $ is the demodulation mapping of the received wideband signal at terminal $q$.

Considering the spatial correlation among the metasurface EM units, We model the channel $\mathbf{G}_{pq}\in \mathbb{C} ^{N_q\times M_p}$ from the TX-SIM at terminal $p$ to the RX-SIM at terminal $q$ as~\cite{9714406}
\begin{align*}
\label{eq8}
\mathbf{G}_{pq}=\left( \mathbf{R}_{p}^{\mathrm{RX}} \right) ^{1/2}\tilde{\mathbf{G}}_{pq}\left( \mathbf{R}_{p}^{\mathrm{TX}} \right) ^{1/2},\tag{9}
\end{align*}
where $\tilde{\mathbf{G}}_{pq}\in \mathbb{C} ^{N_q\times M_p}$ denotes the independent and identically distributed Rayleigh fading channel. $\mathbf{R}_{p}^{\mathrm{TX}}\in \mathbb{C} ^{M_p\times M_p}$ and $\mathbf{R}_{p}^{\mathrm{RX}}\in \mathbb{C} ^{N_p\times N_p}$ represent the spatial correlation matrix at the TX-SIM and that at the RX-SIM, respectively. By considering far-field propagation in an isotropic scattering environment~\cite{9110848}, the spatial correlation matrix can be expressed by~\cite{9716880}
\begin{align*}
\label{eq11}
&[\mathbf{R}_{p}^{\mathrm{TX}}]_{m,\tilde{m}}=\sin\mathrm{c(}2r_{m,\tilde{m}}/\lambda ),\ \tilde{m}\in \mathcal{M} _p,\ m\in \mathcal{M} _p,\tag{10} \\
\label{eq12}
&[\mathbf{R}_{p}^{\mathrm{RX}}]_{\tilde{n},n}=\sin\mathrm{c(}2r_{\tilde{n},n}/\lambda ),\ n\in \mathcal{N} _p,\ \tilde{n}\in \mathcal{N} _p
,\tag{11}
\end{align*}
where $r_{m,\tilde{m}}$ and $r_{\tilde{n},n}$ represent the distance between different corresponding EM units on a single-layer metasurface. The path loss between the transmitter and the receiver is modeled by~\cite{7109864}
\begin{align*}
\label{eq29}
\text{PL}(D) = \text{PL}(D_0) + 10b \log_{10} \left(\frac{D}{D_0}\right) + X_\delta,\tag{12}
\end{align*}
where $\text{PL}(D_0) = 20 \log_{10} \left(\frac{4\pi D_0}{\lambda}\right) \text{ dB}$ is the free space path loss at the reference distance $D_0$, $b$ represents the path loss exponent, $X_\delta$ is a zero mean Gaussian random variable with a standard deviation $\delta$, characterizing the large-scale signal fluctuations of shadow fading.

\begin{figure*}[!t]
\centering
\includegraphics[width=7in]{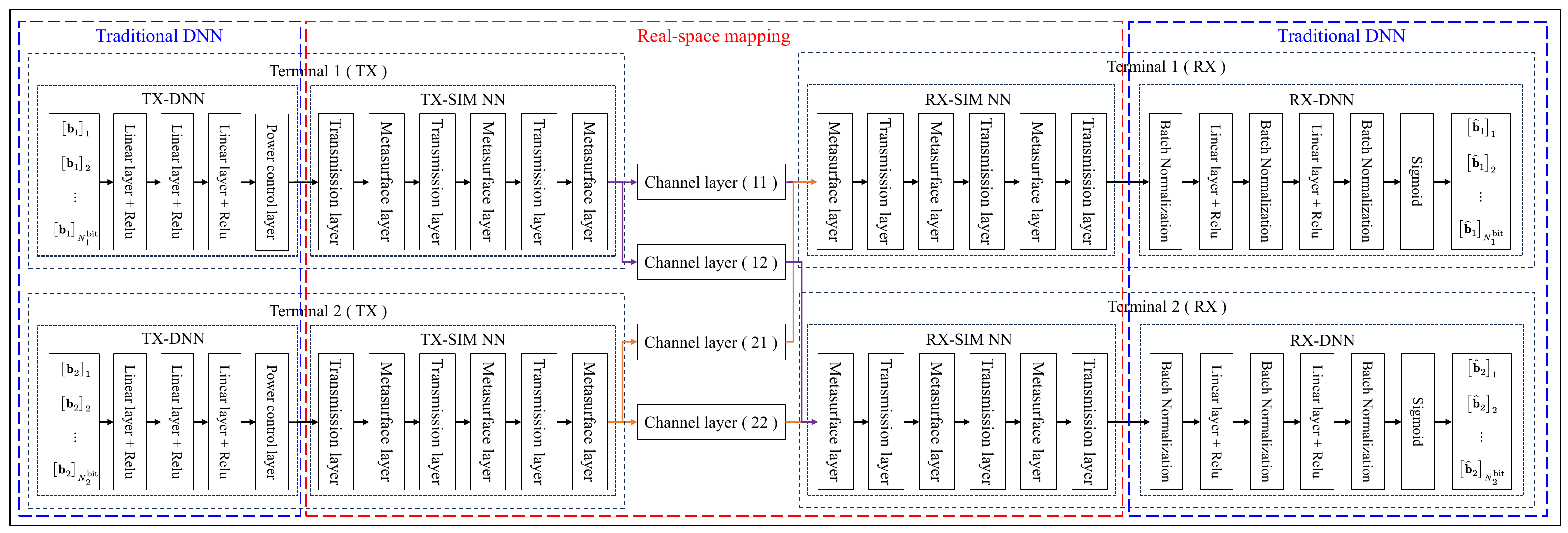}
\caption{Schematic diagram of the EMNN structure.}
\label{Fig4}
\vspace{-0.4cm}
\end{figure*}

\section{E2E problem-solving.} \label{Ⅲ}
\subsection{Problem posing.} 
We define the transmit bit vector and the received bit vector of the E2E CCFD system as
$\mathbf{b}=\left[ \mathbf{b}_{1}^{\mathrm{T}},\mathbf{b}_{2}^{\mathrm{T}} \right] ^{\mathrm{T}}$
and $\hat{\mathbf{b}}=\left[ \hat{\mathbf{b}}_{2}^{\mathrm{T}},\hat{\mathbf{b}}_{1}^{\mathrm{T}} \right] ^{\mathrm{T}}$, respectively. Our goal is to recover the transmitted vector $\mathbf{b}$ from the received vector $\hat{\mathbf{b}}$ as accurately as possible. For this purpose, SIM proactively adjusts the phases of the EM units on the metasurface to synthesize a matched channel, thereby suppressing bit errors caused by SI. Specifically, the objective function is to minimize the bit error rate (BER) between $\mathbf{b}$ and $\hat{\mathbf{b}}$, which is expressed as
\begin{align*}
\label{eq13}
\small{\mathrm{BER}\left( \mathbf{b},\hat{\mathbf{b}} \right) =\frac{\sum_{b=1}^{N_{1}^{\mathrm{bit}}+N_{2}^{\mathrm{bit}}}{P\left\{ \left[ \mathbf{b} \right] _b\ne \left[ \hat{\mathbf{b}} \right] _b \right\}}}{N_{1}^{\mathrm{bit}}+N_{2}^{\mathrm{bit}}}}
.\tag{13}
\end{align*}
Subsequently, the optimization problem for the SIM-assisted E2E CCFD system is formulated as follows
\begin{align*}
\mathcal{P} 1:&\min_{\mathbf{\Phi }_{q}^{l},\mathbf{\Psi }_{q}^{k},\mathbf{p}_q,\mathbf{x}_q} \mathrm{BER}\left( \mathbf{b},\hat{\mathbf{b}} \right),\tag{14a}
\\
&\mathrm{s}.\mathrm{t}.\left\| \mathbf{p}_1 \right\| _1+\left\| \mathbf{p}_2 \right\| _1=P^{\mathrm{t}},\tag{14b}
\\
&\mathbb{E} \left\{ \mathbf{x}_q\left( \mathbf{x}_q \right) ^{\mathrm{H}} \right\} =\mathbf{I}_{A_{q}^{\mathrm{t}}},\tag{14c}
\\
&\left( 1-13 \right),\ q\in \left\{ 1,2 \right\}, \tag{14d}
\end{align*}
where $P^{\mathrm{t}}$ denotes the total transmit power of the system. The search for globally optimal solutions to problem $\mathcal{P}1$ is highly challenging, primarily due to the strong interdependence among the optimization variables and the non-convex unit-modulus constraints imposed on the transmission coefficients of the metasurface EM units. Interestingly, the investigated E2E communication system shares a structural analogy with an autoencoder (AE). Inspired by this insight, we utilize deep learning strategies to train the system, enabling adaptable modulation and demodulation, while concurrently developing an autonomous control framework for the SIM.

\begin{table}[!t] 
    \centering
    \scriptsize
    \caption{Output dimensions of EMNN (All the complex numbers are split into their real and imaginary parts for calculation).}
    \label{table1}
    \renewcommand{\arraystretch}{0.85} 
    \begin{tabular}{cccc} 
        \toprule
                \textbf{Module} & \textbf{Layer} & \textbf{Terminal 1} & \textbf{Terminal 2} \\ 
        \midrule
                & Input & $N_{1}^{\mathrm{bit}}$ & $N_{2}^{\mathrm{bit}}$\\
          & Liner layer + Relu & $N_{1}^{\mathrm{bit}}$ & $N_{2}^{\mathrm{bit}}$\\
         TX-DNN       & Liner layer + Relu & $N_{1}^{\mathrm{bit}}$ & $N_{2}^{\mathrm{bit}}$\\
                & Liner layer + Relu & $2A_{1}^{\mathrm{t}}$ & $2A_{2}^{\mathrm{t}}$\\
                & Power control layer & $2A_{1}^{\mathrm{t}}$ & $2A_{2}^{\mathrm{t}}$\\
        \midrule
                & Transmission layer $1$ & $2M_1$ & $2M_2$\\
                & Metasurface layer $1$ & $2M_1$ & $2M_2$\\
                & Transmission layer $2$ & $2M_1$ & $2M_2$\\
TX-SIM          & Metasurface layer $2$ & $2M_1$ & $2M_2$\\
NN                & ... & ... & ...\\
                & ... & ... & ...\\
                & Transmission layer $L$ & $2M_1$ & $2M_2$\\
                & Metasurface layer $L$ & $2M_1$ & $2M_2$\\
        \midrule
        Channel & Channel layer & $2N_1$ & $2N_2$\\   
        \midrule
                & Metasurface layer $K$ & $2N_1$ & $2N_2$\\
                & Transmission layer $K$ & $2N_1$ & $2N_2$\\
                & Metasurface layer $K-1$ & $2N_1$ & $2N_2$\\
RX-SIM          & Transmission layer $K-1$ & $2N_1$ & $2N_2$\\
NN                & ... & ... & ...\\
              & ... & ... & ...\\
                & Metasurface  layer $1$ & $2N_1$ & $2N_2$\\
                & Transmission layer $1$ & $2A_{1}^{\mathrm{r}}$ & $2A_{2}^{\mathrm{r}}$\\
                
        \midrule
                & Batch Normalization & $2A_{1}^{\mathrm{r}}$ & $2A_{2}^{\mathrm{r}}$\\
                & Liner layer + Relu & $N_{1}^{\mathrm{bit}}$ & $N_{2}^{\mathrm{bit}}$\\
          & Batch Normalization & $N_{1}^{\mathrm{bit}}$ & $N_{2}^{\mathrm{bit}}$\\
         RX-DNN       & Liner layer + Relu & $N_{1}^{\mathrm{bit}}$ & $N_{2}^{\mathrm{bit}}$\\
                & Batch Normalization & $N_{1}^{\mathrm{bit}}$ & $N_{2}^{\mathrm{bit}}$\\
                & Sigmoid & $N_{1}^{\mathrm{bit}}$ & $N_{2}^{\mathrm{bit}}$\\
                & Output & $N_{1}^{\mathrm{bit}}$ & $N_{2}^{\mathrm{bit}}$\\
        \bottomrule
    \end{tabular}
\end{table}

\subsection{Design of EMNN and Model training.}
As illustrated in Fig.~\ref{Fig3} and Fig.~\ref{Fig4}, the fundamental concept of the proposed EMNN is to represent the metasurface layer within the SIM as multiple trainable hidden layers, while describing the practical EM wave propagation process through the forward propagation of a deep neural network. A set of transmit bit vectors $\mathbf{b}$ is randomly generated, and the EMNN is optimized using the mini-batch stochastic gradient descent (MBSGD) algorithm. After training, the learned weights of the hidden layers are extracted and mapped onto the SIM hardware to accurately control each EM unit. In essence, this framework realizes an artificial intelligence (AI)-enabled control paradigm. Additionally, conventional DNNs are employed both between the transmit bit vectors and the transmitted waveforms, and between the received signals and the reconstructed bit vectors, thereby achieving automatic modulation and demodulation.

Considering that signals are typically represented in the complex domain, whereas neural network computations are generally performed in the real domain, the real and imaginary parts of the transmit and receive signals for an antenna are mapped to two separate weights in the hidden layer. For an EM unit, since its magnitude is always unity, only a bijective mapping between its phase coefficient and the corresponding hidden-layer weight is required. The EMNN can be divided into four sub-networks and one channel computation layer, which will be introduced in detail in the following subsections. The parameter configuration of the EMNN is summarized in Table~\ref{table1}.

\textbf{Sub-network 1. TX-DNN.} 
This conventional DNN is deployed at the transmitter side to perform the mapping from the transmit bit vector $\mathbf{b}$ to the transmitted signal $\mathbf{p}_1\odot \mathbf{x}_1$ or $\mathbf{p}_2\odot \mathbf{x}_2$. The DNN comprises multiple linear hidden layers, each utilizing the ReLU activation function. In addition, the power control layer serves as a fixed computational block that adjusts the signal amplitude proportionally to the specified total transmit power $P^{\mathrm{t}}$.

\textbf{Sub-network 2. TX-SIM NN.} 
This fully-connected neural network is specifically designed to establish a one-to-one mapping with the practical TX-SIM hardware. The architecture consists of interleaved transmission and metasurface layers, both derived from the system model abstraction. The transmission layers are directly obtained using~\eqref{eq3} and~\eqref{eq4}, functioning as fixed, non-trainable computational blocks. In contrast, the metasurface layers correspond individually to the actual metasurfaces within the TX-SIM device and are realized as trainable hidden layers.

\textbf{Channel layer.} 
The channel layer serves as a static computational component derived from real-world data through methods like channel estimation or prediction. To account for temporal variations in channel characteristics, we construct two forms of channel computation layers: the statistical channel and the instantaneous channel. Subsequently, transfer learning is applied to optimize the overall model.

\textbf{Sub-network 3. RX-SIM NN.} 
Similar to the design of the TX-SIM NN, this is also a specialized fully-connected neural network aimed at establishing a bijective relationship with the actual RX-SIM hardware. The design principles of the transmission layers and metasurface layers are identical to 
those described earlier and are therefore omitted here.

\begin{table}[!t] 
    \centering
    \caption{Simulation parameter settings}
    \label{table2}
    \scriptsize
    \renewcommand{\arraystretch}{0.85} 
    \begin{tabular}{ll} 
        \toprule
                \textbf{System parameters} & \textbf{Value} \\
        \midrule
                Frequency ($f$) & 28 GHz \\
                Wavelength ($\lambda$) & 10.7 mm \\
                Distance between Terminal 1 and Terminal 2 ($D$) & 50 m \\
                Number of bits transmitted by Terminal 1 ($N_{1}^{\mathrm{bit}}$) & 12\\
                Number of bits transmitted by Terminal 2 ($N_{2}^{\mathrm{bit}}$) & 8\\
                Monte carlo & 100 \\
        \toprule
                \textbf{Channel parameters} & \textbf{Value} \\
        \midrule
                Path loss reference distance ($D_0$) & 1 m \\
                Path loss exponent ($b$) & 3.5 \\
                Path loss shadowing fading variance ($\delta$) & 9 dB \\
                Receiver noise ($\sigma^2$) & -110 dBm \\
        \toprule
                \textbf{SIM parameters} & \textbf{Value} \\
        \midrule
                Number of layers of terminal 1 TX-SIM ($L_1$) & 3 \\
                Number of layers of terminal 1 RX-SIM ($K_1$) & 3 \\
                Number of layers of terminal 2 TX-SIM ($L_2$) & 3 \\
                Number of layers of terminal 2 RX-SIM ($K_2$) & 3 \\
                Number of EM units of terminal 1 TX-SIM  & $81=9\times 9$ \\
                Number of EM units of terminal 1 RX-SIM  & $81=9\times 9$ \\
                Number of EM units of terminal 2 TX-SIM  & $81=9\times 9$ \\
                Number of EM units of terminal 2 RX-SIM  & $81=9\times 9$ \\
                Number of terminal 1 TX antennas & $16=4\times 4$ \\
                Number of terminal 1 RX antennas & $16=4\times 4$ \\
                Number of terminal 2 TX antennas & $9=3\times 3$ \\
                Number of terminal 2 RX antennas & $9=3\times 3$ \\
                Spacing of EM units in SIM ($d$) & $\lambda/2$ \\
                SIM layer spacing ($r$) & $\lambda/2$ \\
        \toprule
                \textbf{Training and testing parameters} & \textbf{Value}\\ 
        \midrule
                Loss function & BCE\\
                Initialization & Xavier\\
                Optimizer & AdamW\\
                Training epoch ($E$)  & 2000\\
                Learning rate & $0.005$\\
                Batch size ($N^{\mathrm{batch}}$) & 1000\\
                Performance metric & BER\\
                Test scale & 100000\\
                Learning rate decay & 0.95\\
        \bottomrule
        
    \end{tabular}
\end{table}

\textbf{Sub-network 4. RX-DNN.} 
Similar to the TX-DNN, this is also a conventional DNN. The DNN alternates between batch normalization and linear layers, with all linear layers employing the ReLU activation function. Finally, a sigmoid function is applied to normalize the output to the range $\left(0,1\right)$, followed by hard decision to recover the bit vector.

For accurately recovering the original bit vector in time-varying channels, the E2E system needs to be retrained periodically based on CSI information. To accelerate this process, we can employ transfer learning. Specifically, a base model is first trained using the statistical CSI, and then the model is fine-tuned in real time according to the instantaneous CSI. The loss function used during the model training is the binary cross-entropy (BCE), which can be computed as 
\begin{align*}
\label{eq15}
\mathcal{L} \left( \mathbf{b},\hat{\mathbf{b}} \right) =&-\frac{1}{N^{\mathrm{batch}}}\sum_{\mathfrak{b} =1}^{N^{\mathrm{batch}}}{\sum_{b=1}^{N_{1}^{\mathrm{bit}}+N_{2}^{\mathrm{bit}}}{\left\{ \left[ \mathbf{b}_{\mathfrak{b}} \right] _b\log \left( \left[ \hat{\mathbf{b}}_{\mathfrak{b}} \right] _b \right) \right.}}
\\
&\left. +(1-\left[ \mathbf{b}_{\mathfrak{b}} \right] _b)\log \left( 1-\left[ \hat{\mathbf{b}}_{\mathfrak{b}} \right] _b \right) \right\},\tag{15}
\end{align*}
where $N^{\mathrm{batch}}$ denotes the training batch size. Similar to~\eqref{eq13}, BCE is employed to evaluate the consistency between the input and the output. 


\begin{figure}[!t]
\centering
\includegraphics[width=2.6in]{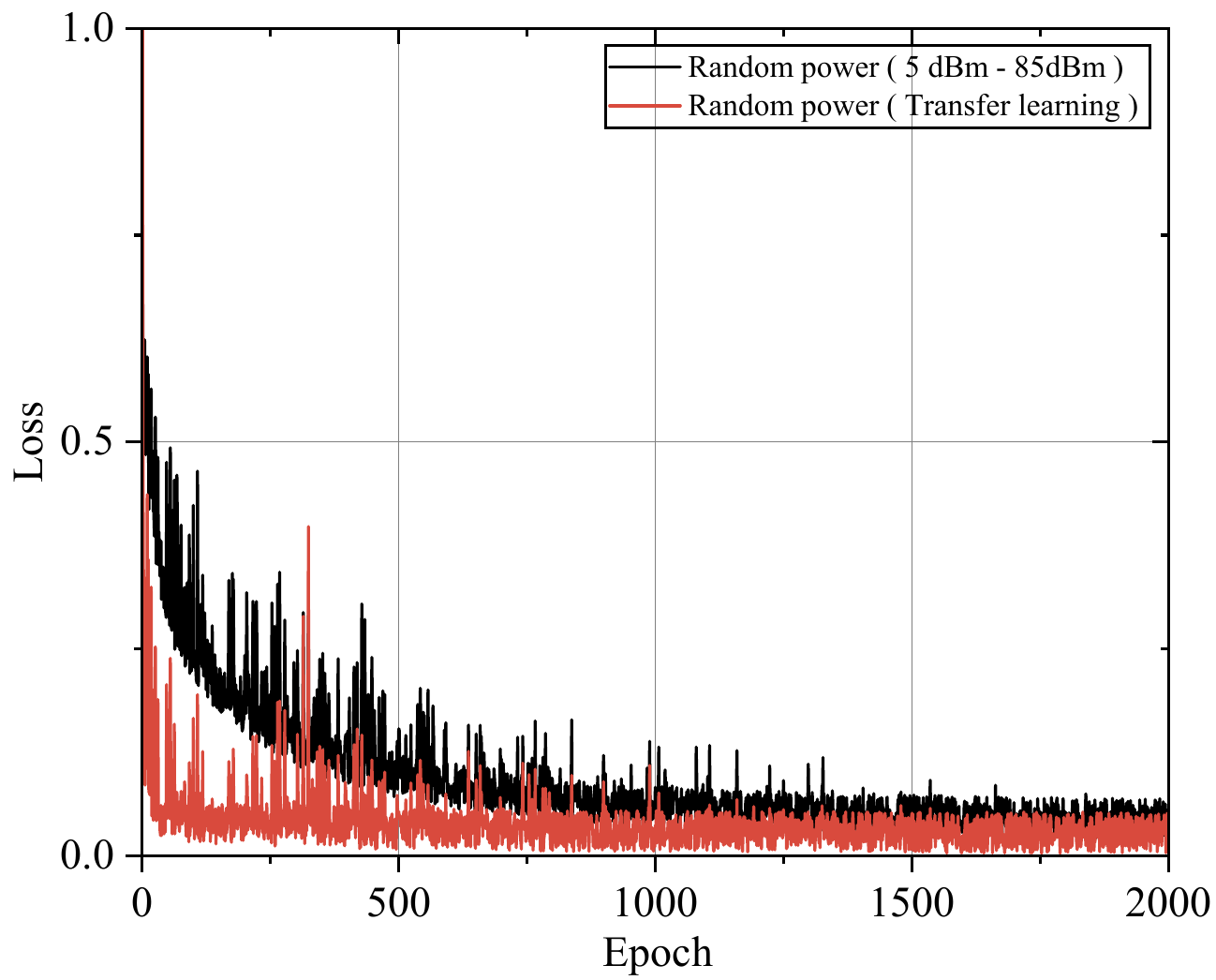}
\caption{Iteration graph of EMNN training loss.}
\label{Fig5}
\vspace{-0.3cm}
\end{figure}

\begin{figure}[!t]
\centering
\includegraphics[width=2.6in]{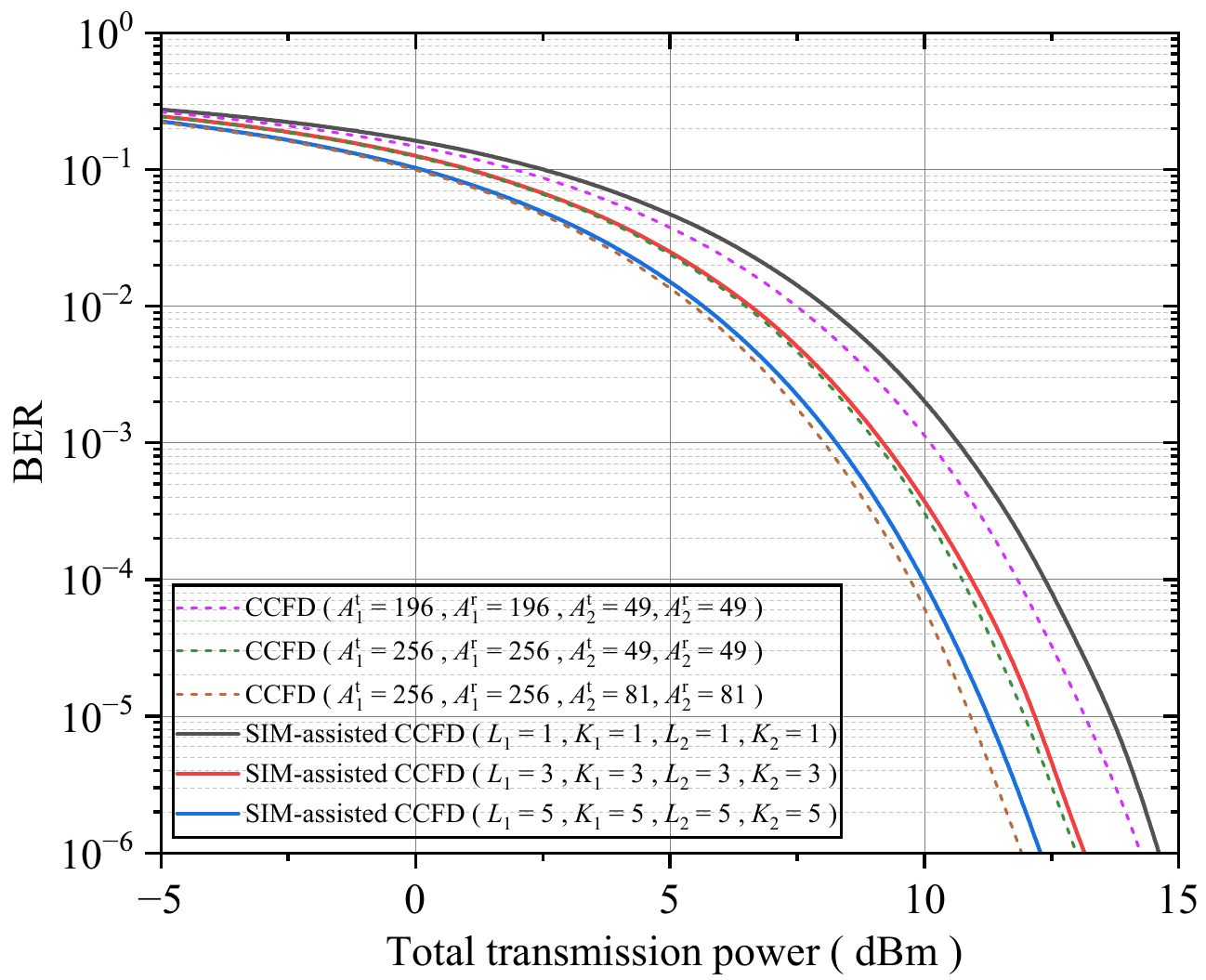}
\caption{Performance comparison of different numbers of metasurface layers.}
\label{Fig6}
\vspace{-0.1cm}
\end{figure}

\section{Simulation Results.} \label{Ⅳ}

Given that the framework illustrated in Fig.~\ref{Fig4} involves intricate cross-layer modeling and system-level implementation, this section concentrates on verifying its key component—the EMNN approach trained through transfer learning within SIM-assisted E2E CCFD systems. The effectiveness of this method is crucial to the overall performance of the entire framework; thus, its verification outcomes largely indicate the practicality of the proposed design. In particular, a quasi-static block-fading channel is considered. Initially, the EMNN is trained using statistical CSI to develop a baseline model. Subsequently, several sets of instantaneous channel realizations are randomly generated, each used to update the EMNN by substituting the channel layer and performing fine-tuning via transfer learning. Finally, the system performance is evaluated by averaging the testing results obtained under various channel conditions. The system simulation parameters employed are listed in Table~\ref{table2}. Unless otherwise noted, all subsequent simulation outcomes are derived based on these settings.

\begin{figure}[!t]
\centering
\includegraphics[width=2.6in]{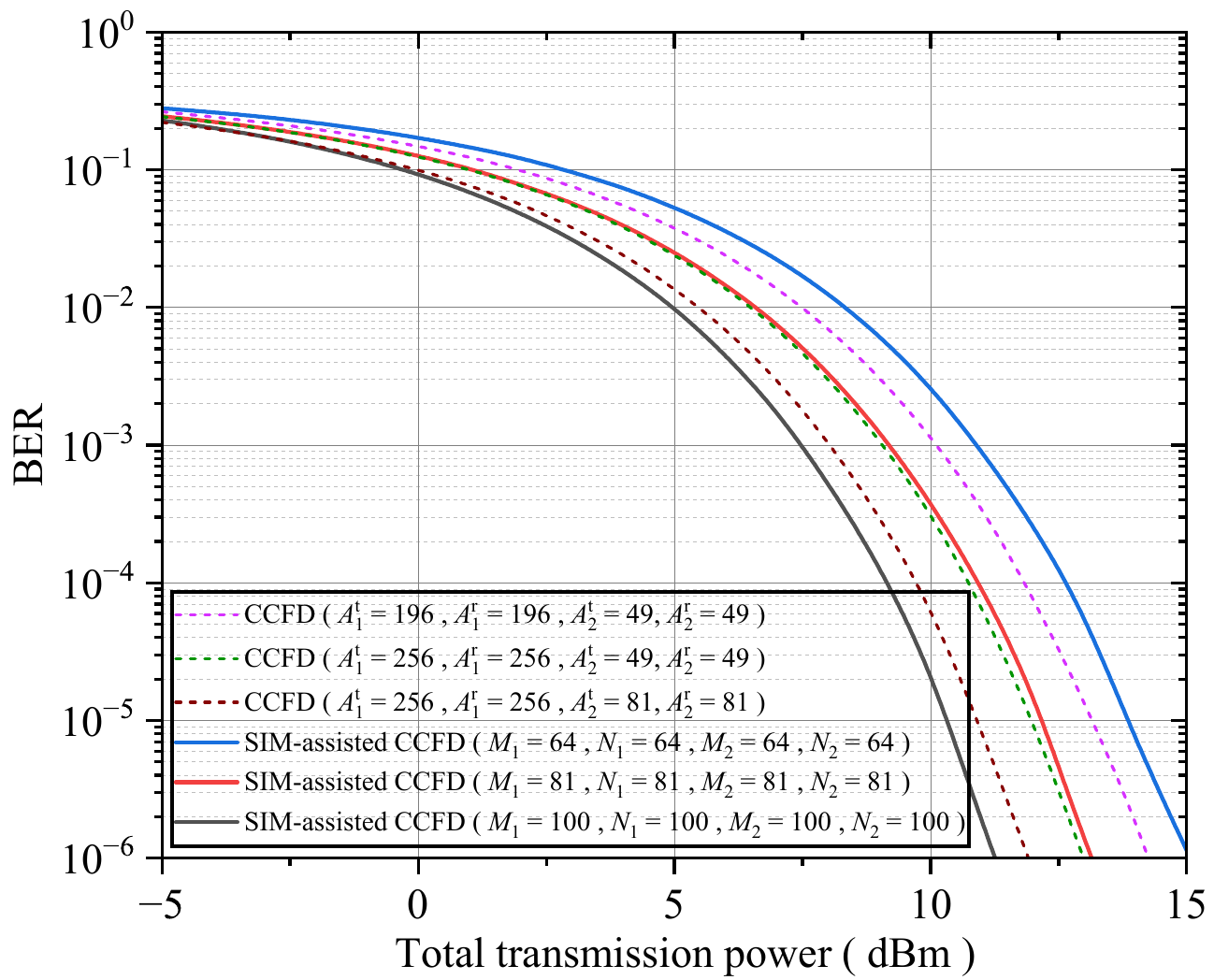}
\caption{Performance comparison of different numbers of EM unit quantities.}
\label{Fig7}
\vspace{-0.1cm}
\end{figure}

\begin{figure}[!t]
\centering
\includegraphics[width=2.6in]{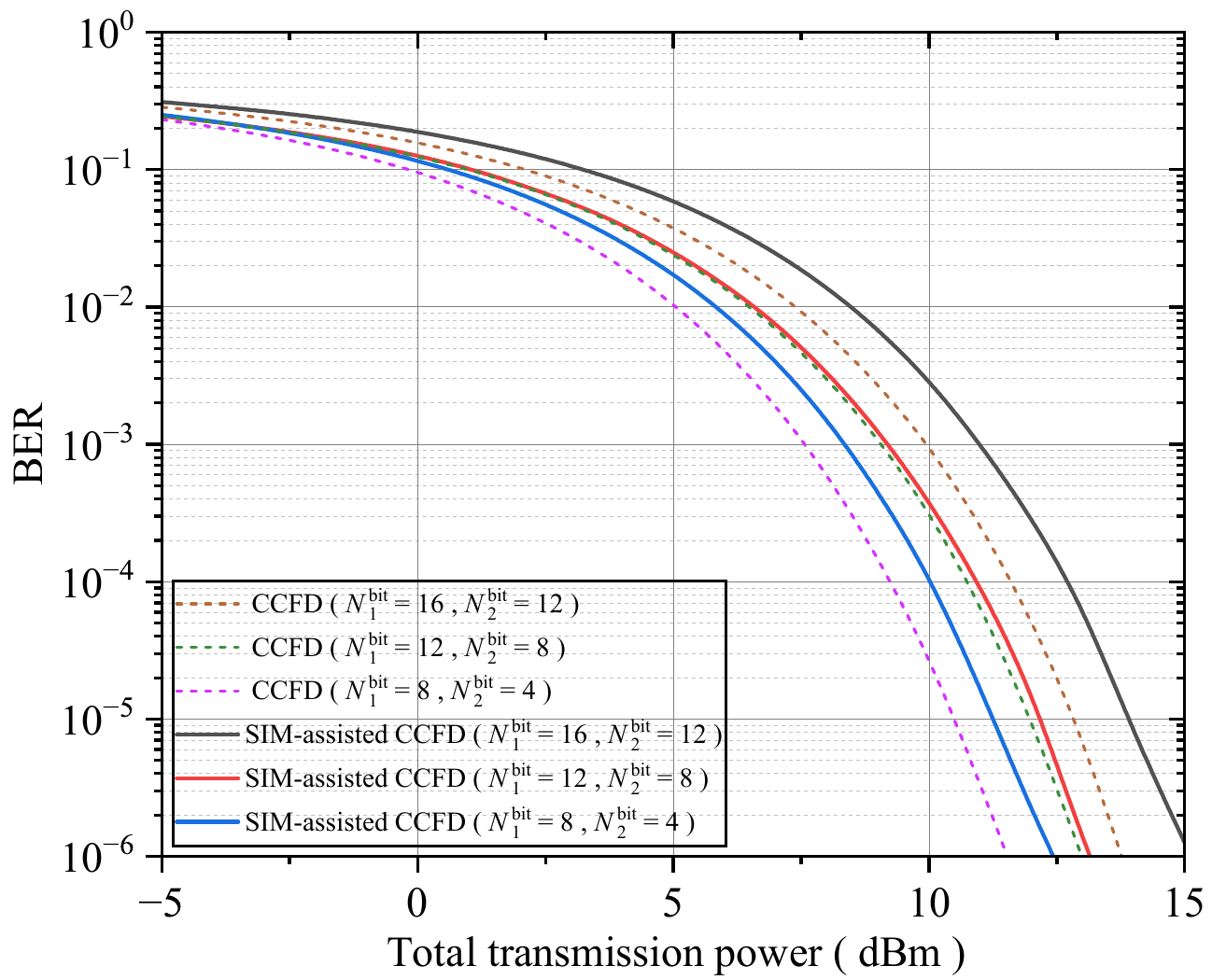}
\caption{Performance comparison of different transmission bit numbers.}
\label{Fig8}
\vspace{-0.1cm}
\end{figure}

Fig.~\ref{Fig5} shows the iteration curves of our EMNN training accelerated by transfer learning. To achieve good performance under both high and low transmission power conditions, we use the Beta distribution to randomly generate training transmission power. It is observed that the transfer learning-based training method can achieve performance comparable to direct training while significantly accelerating the training process. Fig.~\ref{Fig6} and Fig.~\ref{Fig7} show the impact of the number of metasurface layers and the scale of EM units on the BER performance of the proposed SIM-assisted CCFD system. The results indicate that increasing the number of metasurface layers and the scale of EM units enhances the signal processing capability of the SIM, thereby effectively suppressing SI and improving the overall system performance. In particular, it is observed that by integrating the SIM device, a CCFD system configured as (16T1 16R1 9T2 9R2) can achieve performance comparable to a traditional CCFD system configured as (256T1 256R1 49T2 49R2). This validates the feasibility of leveraging integrated SIM devices to assist SI suppression in the wave domain, thereby further 
exploiting the potential of spectrum resources. Fig.~\ref{Fig8} shows the BER curves for different numbers of transmitted bits within one symbol period. It can be observed that when the number of bits per symbol in the CCFD system decreases, the system BER decreases accordingly.


\section{Conclusion.} \label{Ⅴ}
To more effectively suppress SI in the CCFD system, we propose integrating a SIM into the RF front end to assist SI mitigation. By leveraging the powerful signal processing capability of SIM in the wave domain, the CCFD system can fully exploit spectral resources to enhance overall system performance. Furthermore, we design an EMNN network for the SIM-assisted E2E CCFD system and implement full-link joint optimization based on transfer learning. Simulation results demonstrate that, under complex channel conditions, the SIM-assisted E2E CCFD system achieves more robust bitstream transmission compared with the conventional CCFD system. This study highlights the great potential of the EMNN and SIM-assisted E2E CCFD system in the design of next-generation intelligent transceivers.

\bibliographystyle{IEEEtran}
\bibliography{references}

\end{document}